\documentclass[usenatbib]{mnras}

\usepackage[T1]{fontenc}
\usepackage{ae,aecompl}

\usepackage{graphicx}      
\usepackage{epstopdf}
\usepackage{amsmath}    
\usepackage{amssymb}    
\usepackage{booktabs}
\usepackage{bm}

\newcommand{\kB}{k_\mathrm{B}}
\newcommand{\beq}{\begin{equation}}
\newcommand{\eeq}{\end{equation}}
\newcommand{\dd}{\mathrm{d}}
\newcommand{\Ts}{T_\mathrm{s}}
\newcommand{\Tb}{T_\mathrm{b}}
\newcommand{\rhob}{\rho_\mathrm{b}}
\newcommand{\gcc}{\mbox{ g~cm$^{-3}$}}
\newcommand{\Msun}{\mathrm M\odot}
\newcommand{\e}{\mathrm{e}}

\title[Diffusive heat blankets of neutron stars]{
Diffusive heat blanketing envelopes of neutron stars}

\author[M. V. Beznogov,  A. Y. Potekhin, D. G. Yakovlev]{
M. V. Beznogov$^{1}$\thanks{E-mail: \href{mailto:mikavb89@gmail.com}{mikavb89@gmail.com}},
A. Y. Potekhin$^{2,3,4}$, D. G. Yakovlev$^{2}$
\\
$^{1}$St.~Petersburg Academic University, 8/3 Khlopina st.,
St.~Petersburg 194021, Russia \\
$^{2}$Ioffe Institute, 26 Politekhnicheskaya st.,
St.~Petersburg 194021, Russia \\
$^{3}$Central Astronomical Observatory at Pulkovo, Pulkovskoe Shosse 65,
Saint Petersburg 196140, Russia \\
$^{4}$Saint-Petersburg Polytechnic University,
     29 Politekhnicheskaya st., Saint Petersburg 195251,
       Russia}

\date{Accepted . Received ; in original form}
\pubyear{2016}

\begin{document}
\label{firstpage}
\pagerange{\pageref{firstpage}--\pageref{lastpage}}
\maketitle

\begin{abstract}
We construct new models of outer heat blanketing envelopes of
neutron stars composed of binary ion mixtures (H -- He, He -- C, C
-- Fe) in and out of diffusive equilibrium. To this aim, we
generalize our previous work on diffusion of ions in isothermal
gaseous or Coulomb liquid plasmas to handle non-isothermal systems.
We calculate the relations between the effective surface temperature
$\Ts$ and the temperature $\Tb$ at the bottom of heat blanketing
envelopes (at a density $\rhob \sim 10^8-10^{10}$ \gcc) for
diffusively equilibrated and non-equilibrated distributions of ion
species at different masses $\Delta M$ of lighter ions in the
envelope. Our principal result is that the $\Ts - \Tb$ relations are
fairly insensitive to detailed distribution of ion fractions over
the envelope (diffusively equilibrated or not) and depend almost
solely on $\Delta M$. The obtained relations are approximated by
analytic expressions which are convenient for modeling the evolution
of neutron stars.
\end{abstract}


\section{Introduction}
\label{sec:intro}

It is well known (see, e.g., \citealt*{YP04,PPP15}, and
references therein) that modeling thermal evolution of neutron stars
and comparing the results with observations gives an important method
to explore the properties of superdense matter in neutron star cores.
As a rule, such studies require theoretical determination of internal
temperatures of neutron stars from their observable surface
temperatures $\Ts$. The internal temperatures are typically much
higher than $\Ts$ because neutron stars possess thin surface heat
blanketing envelopes with poor thermal conduction. They produce good
thermal insulation for stellar interiors.

The composition of these envelopes is a priory unknown; they may
contain heavy (iron-like) elements or some amount of lighter (for
instance, accreted) elements. The composition affects the insulation
and introduces significant uncertainties in the studies of internal
structure of neutron stars (e.g., \citealt{Weisskopf_etal11}).
The situation looks funny. The properties of the heat blanketing
envelopes are determined by the physics of
ordinary plasma, which is  much more elaborated
 than the largely unknown physics of dense neutron star interiors
(e.g., \citealt*{HPY07,L14}, and references therein). Nevertheless,
the uncertainties in our knowledge of the chemical composition of
the heat blanketing envelopes greatly complicate the investigation
of mysterious neutron star interiors. This motivates further study
of the envelopes with different chemical composition.

It is our aim to develop new models of the heat blanketing envelopes.
Formally, these envelopes extend from the bottom of the stellar
atmosphere to some density $\rho=\rhob\sim 10^8-10^{10}$ \gcc{} which
can be chosen differently depending on a specific problem (Sect.{}
\ref{sec:ModelParam}). The temperature $\Tb$ at the bottom of the heat
blanket ($\rho=\rhob$) depends on $\Ts$, so that the main problem of
practical interest is to obtain the $\Ts - \Tb$ relation. This
relation can be further used as a boundary condition for calculating
the temperature distribution $T(\bm{r},t)$ within the star at $\rho>
\rhob$ (e.g., \citealt{YP04,PPP15}, and references therein).

The heat blanketing envelopes are geometrically thin (their typical
depth does not exceed a few hundreds meters) and contain a very
small mass $ \lesssim 10^{-7}\, \Msun$. Therefore, a small local
part of the envelope can be approximated by a plane-parallel layer
in a locally flat geometry with a constant surface gravity
$g_\mathrm{s}$ (e.g., \citealt*{GPE83}). One usually assumes
hydrostatic equilibrium, quasi-stationary approximation, and a
locally constant thermal flux which emerges from the stellar
interior to the surface. Here we adopt these standard assumptions
which allow us to perform a relatively easy one-dimensional
calculation of the $\Ts-\Tb$ relation in a local part of the
surface. Physical conditions can vary over the entire surface (e.g.,
due to the presence of a strong magnetic field, -- see
\citealt{PPP15} and references therein); then the $\Ts-\Tb$ relation
will also vary.

The $\Ts-\Tb$ relations have been calculated in many publications. Let
us mention the pioneering work by \citet{GPE83} who considered the
envelopes made of iron. \citet*{PCY97} studied the heat blankets which
contain either iron or successive layers of hydrogen, helium, carbon,
and iron. In the latter case the density and temperature ranges for
the existence of any element have been restricted by the conditions of
nuclear transformations (nuclear reactions and beta captures) and the
total mass $\Delta M$ of light elements (H, He and C) has been treated
as a free parameter. Similar envelopes composed of carbon (of mass
$\Delta M$) on top of iron have been constructed by
\citet{Yakovlev_etal11}. \citet{PYCG03} generalized the results of
\citet{PCY97} to the case of strong magnetic fields. In the presence
of very strong (magnetar's) fields in hot neutron star envelopes the
structure of heat blanketing layers can be affected by neutrino
emission (\citealt*{PCY07}; \citealt{KPYC09}). In such a case, the heat
flux through the envelope is not constant. Therefore, the $\Ts-\Tb$
relation does not produce
a proper boundary condition for the neutron star cooling problem;
it should be replaced by a $F_\mathrm{b}-\Tb$
relation, where $F_\mathrm{b}$ is the radial heat flux density at
$\rho=\rhob$. We will not consider the latter case in the present
paper.

All these studies have assumed the presence of only one ion
(nucleus) species at any density and temperature in the heat
blanketing envelope. Here we neglect the effects of magnetic fields
but consider the envelopes containing mixtures of ion species. The
envelopes containing ion mixtures have been studied earlier (e.g.,
\citealt*{HameuryHB83,DeBlasio00,CB03,CB04,CB10}). For example,
\citet{CB03,CB04} and \citet{CB10} have focused on diffusive nuclear
burning of a small amount of lighter elements which diffuse in
deeper layers. The authors have assumed diffusive equilibrium but
neglected the effects of temperature gradients on Coulomb terms (see
also Sects.{} \ref{sec:Envelopes} and \ref{sec:EquilibResults}). We
will consider the diffusive equilibrium including temperature
gradients. We will study also ion distributions out of diffusive
equilibrium, but we neglect the effects of diffusive nuclear
burning.

Diffusion in ion mixtures is a complicated problem.
We focus on the diffusion in dense stellar plasmas where the ions can
be moderately or strongly coupled by Coulomb forces. Such plasmas are
characteristic for white dwarfs and the envelopes of neutron stars.

Consider a non-magnetized multicomponent plasma consisting of
several ion species ($\alpha=j$, $j=1,2,\ldots$) and neutralizing
electron background ($\alpha=\e$).  Let $A_j$ and $Z_j$ be the mass
and charge numbers of ion species $j$, and $n_\alpha$ be the number
density of particles $\alpha$,  with
\begin{equation}
   n_\e=\sum_j Z_j n_j
\label{e:Neutrality}
\end{equation}
due to electric neutrality. It is convenient to introduce (cf.
\citealt{HPY07}) the average Coulomb coupling parameter
$\overline{\Gamma}=\Gamma_0
\overline{Z^{{5}/{3}}}\,\overline{Z}^{{1}/{3}}$, where the average
value of any quantity $f$ is defined as
$\overline{f}\equiv\sum_j {x_j}f_j$,
$x_j={n_j}/{n}$ is a number fraction of the ion species $j$,
$n=\sum_j n_j$ is the total number density of the ions,
$\Gamma_0={e^2}/(a \kB T)$,  $e$ is the elementary charge,
$a=\left(4\pi n/3\right)^{-{1}/{3}}$ is the ion sphere radius, $\kB$
is the Boltzmann constant and $T$ is the temperature. If
$\overline{\Gamma}\gg 1$ the ions are strongly coupled (highly
non-ideal), whereas at $\overline{\Gamma}\ll 1$ they are weakly
coupled; $\overline{\Gamma}\sim 1$ refers to the intermediate
coupling.

We will mostly focus on diffusion-equilibrium heat blanketing envelopes.
Unless stated otherwise, this means the equilibrium with respect to
diffusion as well as overall hydrostatic equilibrium, not the total
thermodynamic equilibrium (obviously, a non-isothermal system cannot
be in the state of total thermodynamic equilibrium).

In Sect.{} \ref{sec:DiffFlux} we present a general formulation of the
diffusion and thermal diffusion problem. In Sects.\
\ref{sec:Envelopes}, \ref{sec:EnvModel}, \ref{sec:ModelParam},
\ref{sec:EquilibResults} we apply this general theory to diffusively
equilibrated heat blanketing envelopes of neutron stars. We will
also study non-equilibrated envelopes (Sect.\
\ref{sec:NonEquilib}) and present analytic fits to our $\Tb-\Ts$
calculations in Appendix~\ref{sec:Fits}.

\section{General expressions for diffusive fluxes}
\label{sec:DiffFlux}

The general idea for deriving diffusive fluxes is the same as
described by \cite{BY13,BY14b} for an isothermal plasma. We start
from generalized thermodynamic forces $\widetilde{\bm{f}}_\alpha$
acting on particles $\alpha$ and take into account a temperature gradient.
Therefore, $\widetilde{\bm{f}}_\alpha$ includes an
additional term proportional to $\bm{\nabla} T$,
\begin{equation}
   \widetilde{\bm{f}}_\alpha = \bm{f}_\alpha -
   \left(\bm{\nabla}\mu_\alpha -\left.\frac{\partial \mu_\alpha }
   {\partial T}\right|_P \bm{\nabla}T\right).
\label{e:GenForce}
\end{equation}
Here $\bm{f}_\alpha$ is a total force, acting on particles $\alpha$,
$\mu_\alpha$ is their chemical potential, and $\bm{\nabla}$ is the gradient
operator in the proper reference frame. For instance, in the spherical
coordinates $(r,\theta,\varphi)$ for a non-rotating star with a
spherically symmetric mechanical structure we have
(cf., e.g., \citealt{HPY07})
\beq
    \bm{\nabla} = \left(\begin{array}{c}
    \mathrm{e}^{-\Lambda(r)} {\partial}/{\partial r}
\\
     r^{-1}\,{\partial}/{\partial\theta}
\\
     (r\sin\theta)^{-1}\,{\partial}/{\partial\varphi}
      \end{array}\right),
\label{e:nabla}
\eeq
where $\Lambda(r) = -(1/2)\ln(1-GM_r/c^2r)$ is the metric function
which determines the space curvature in the radial direction,
$M_r=4\pi\int_0^r\rho(r)r^2\,\dd r$ is the gravitational mass inside a sphere of circumferential
radius $r$, $G$ is the gravitational constant and $c$ is the speed
of light. In heat blanketing envelopes of neutron stars the
hydrostatic balance is mainly controlled by the electric and
gravitational forces. Therefore,
\begin{equation}
   \bm{f}_\alpha = Z_\alpha e \bm{E}+m_\alpha \bm{g},
\label{e:Force}
\end{equation}
where $Z_\alpha e$ and $m_\alpha$ are charge and mass of particles
$\alpha$, respectively ($Z_\mathrm{e}=-1$); $\bm{g}$ is a gravitational
acceleration (defined below) and $\bm{E}$ is an electric field due
to plasma polarization in the external gravitational field.

Deviations from the diffusion equilibrium are characterized by the quantities
$\bm{d}_\alpha$ introduced in the same way as in \cite{BY13,BY14b},
\begin{equation}
   \bm{d}_\alpha = \frac{\rho_\alpha}{\rho}
   \sum_\beta n_\beta \widetilde{\bm{f}}_\beta- n_\alpha
                 \widetilde{\bm{f}}_\alpha,
\label{e:d-vect}
\end{equation}
where $\rho_\alpha = m_\alpha n_\alpha$ is a mass density of particles
$\alpha$ and $\rho$ is the total mass density. Clearly, $\sum_\alpha
\bm{d}_\alpha = 0$. Using equations \eqref{e:GenForce} and \eqref{e:Force},
the Gibbs-Duhem
relation $\sum_\alpha n_\alpha \bm{\nabla}\mu_\alpha = \bm{\nabla}P - S
\bm{\nabla}T$ ($S$ being the entropy density) and the electric neutrality
condition \eqref{e:Neutrality}, we obtain
\begin{equation}
   \sum_\alpha n_\alpha \widetilde{\bm{f}}_\alpha = \rho \bm{g} - \bm{\nabla}P.
\label{e:SumF}
\end{equation}
We are interested in the heat blanketing envelopes
at hydrostatic equilibrium. Then the right-hand side of equation \eqref{e:SumF} is
zero, and equation \eqref{e:d-vect} simplifies to
\begin{equation}
   \bm{d}_\alpha = - n_\alpha \widetilde{\bm{f}}_\alpha.
\label{e:d_short}
\end{equation}
Using equations~\eqref{e:GenForce} and \eqref{e:Force},
equation~\eqref{e:d_short} can be rewritten as
\begin{equation}
   \bm{d}_\alpha = -\frac{\rho_\alpha}{\rho}\bm{\nabla}P
               - Z_\alpha n_\alpha e \bm{E}
    +n_\alpha \left( \bm{\nabla}\mu_\alpha
      - \frac{\partial \mu_\alpha }{\partial T}\bigg|_P \bm{\nabla}T\right).
\label{e:d-alt}
\end{equation}
Since the electrons are much lighter than the ions, we use the adiabatic (or
Born-Oppenheimer) approximation, which assumes the electron quasi-equilibrium
with respect to the motion of atomic nuclei. In this approximation
$\bm{d}_\mathrm{e}=0$ and $m_\mathrm{e} \to 0$,
which leads to $\widetilde{\bm{f}}_\mathrm{e}=0$ and to
\begin{equation}
   e\bm{E} = -\left(\bm{\nabla}\mu_\mathrm{e}
       - \left.\frac{\partial \mu_\mathrm{e}}{\partial T}
            \right|_P \bm{\nabla}T\right).
\label{e:eE}
\end{equation}
This expression can be rewritten in terms of chemical potentials of ions,
using standard thermodynamic relations (e.g., \citealt{LL_Stat93}).

Chemical potentials are usually known as functions of temperature and number
densities. It is, therefore, useful to express $\partial \mu / \partial T$ at
constant $P$ and $x_j$ in terms of $\partial \mu / \partial T$ at constant
$n_j$,
\begin{align}
\begin{split}
   & \left.\frac{\partial \mu}{\partial T}\right|_{P,\{x_j\}} =
       \left.\frac{\partial \mu}{\partial T}\right|_{\{n_j\}}
     - \left.\frac{\partial P}{\partial T}\right|_{\{n_j\}} \\
   & \times \sum_j n_j \frac{\partial \mu}{\partial n_j}\bigg|_{T,\{n_k|k\neq j\}}
       \left(  \sum_j n_j \frac{\partial P}{\partial n_j}\bigg|_{T,\{n_k|k\neq j\}} \right)^{-1}.
\end{split}
\label{e:MuPDeriv}
\end{align}

Phenomenological transport equations for the diffusive fluxes can be written as
\begin{align}
   \bm{J}_\alpha =  \frac{n m_\alpha}{\rho\kB T}
     \sum_{\beta \neq \alpha} m_\beta D_{\alpha\beta}\bm{d}_\beta
       - D_\alpha^{ T} \frac{\bm{\nabla}T}{T},
\label{e:J}
\end{align}
where $D_{\alpha\beta}$ is a generalized diffusion coefficient for particles
$\alpha$ with respect to particles $\beta$, $D_\alpha^{T}$ is a thermal
diffusion coefficient of particles $\alpha$, and the coefficient
before the sum is
chosen so as to match the conventional definition of $D_{\alpha\beta}$
(e.g., \citealt*{Hirsh54,LL_Kin}; cf. \citealt{BY13}).

\section{Theory of heat-blanketing envelopes in diffusive equilibrium}
\label{sec:Envelopes}

Consider  a neutron star outer heat-blanketing envelope composed of a
mixture of two ion species and neutralizing electron background, the so called
binary ionic mixture (BIM). In order to construct the diffusion-equilibrium
envelope, we use several assumptions. First,
electrons have little impact on the transport of ions (see
\citealt{Paquette86}) so that the ion subsystem can be studied
(quasi-)independently. This means that we can set $\bm{J}_\mathrm{e} = 0$ and,
consequently, $\bm{J}_1=-\bm{J}_2$. Second, the thermal diffusion term may
affect the result. However it is usually small compared to
ordinary diffusion which allows us to neglect thermal diffusion (we will
briefly discuss this statement in Sect.{} \ref{sec:NonEquilib}). With these assumptions, one can simplify the diffusive flux of ions,
\begin{equation}
   \bm{J}_2 = -\bm{J}_1=\frac{n m_1 m_2}{\rho \kB T} D_{12} \bm{d}_1,
\label{e:J-BIM}
\end{equation}
where $D_{12}$ is the interdiffusion coefficient. According to
equation \eqref{e:J-BIM}, the diffusion equilibrium $\bm{J}_2 = 0$
is equivalent to the condition $\bm{d}_1 = 0$, or to
$\widetilde{\bm{f}}_1=0$ if we take into account \eqref{e:d_short}.
Equation  $\widetilde{\bm{f}}_1=0$ (along with $\widetilde{\bm{f}}_2=0$ and
$\widetilde{\bm{f}}_\mathrm{e}=0$ as discussed in Sect.\
\ref{sec:DiffFlux}) can then be used to calculate the equilibrium
configuration. Combining equations \eqref{e:GenForce},
\eqref{e:Force} and \eqref{e:MuPDeriv} we obtain the following
system of linear first order differential equations,
\beq
   \widetilde{\bm{\nabla}}\mu_e = -e\bm{E},
\quad
   \widetilde{\bm{\nabla}}\mu_j = m_j \bm{g} + Z_j e\bm{E} ,
\label{e:SysEq}
\eeq
where $ \widetilde{\bm{\nabla}}$ is defined as
\begin{align}
\begin{split}
   \widetilde{\bm{\nabla}}\mu_\alpha \equiv
   \sum_j \frac{\partial \mu_\alpha}{\partial n_j}\, \bm{\nabla} n_j
   &+\frac{\partial P}{\partial T} \sum_j n_j \frac{\partial \mu_\alpha}{\partial n_j} \\
   &\times \left(\sum_k n_k \frac{\partial P}{\partial n_k}\right)^{-1} \bm{\nabla}T.
\end{split}
\label{e:CorrGrad}
\end{align}
Subscripts $j$ and $k$ run over all ion species, $\mu_\alpha$ and
$P$ are assumed to be known together with their derivatives as
functions of $\{n_j\}$ and $T$, and the unknowns are
$\bm{\nabla}n_j$ and $e\bm{E}$. Note that by neglecting the thermal
diffusion term in the diffusive flux \eqref{e:J-BIM}, we have also
excluded the reciprocal Dufour effect in Eq. \eqref{e:HeatEq} [see
below]. In this approximation we do not need an explicit expression
for $D_{\alpha\beta}$. However, generally, taking into account
thermal diffusion, the Dufour effect or transformations of ions
(e.g., because of chemical or nuclear reactions) one needs both the
diffusion and thermal diffusion coefficients to find the equilibrium
configuration.

The closure of the system of equations \eqref{e:SysEq} and \eqref{e:CorrGrad} is
provided by the heat transport equation (see, e.g., \citealt{PPP15} and
references therein)
\begin{equation}
   \mathrm{e}^{-\Phi} \kappa\, \bm{\nabla} \left( T
   \mathrm{e}^{\Phi}\right)  = -\bm{F}_T,
\label{e:HeatEq}
\end{equation}
where $\bm{F}_T$ is a local thermal flux, $\kappa$ is a thermal conductivity,
$\bm{\nabla}$ is given by equation \eqref{e:nabla}, and $\Phi(\bm{r})$ is the
metric function which determines gravitational redshift
(an effective dimensionless gravitational potential).

Since the thickness of the heat blanketing envelope is
much smaller than the (circumferential) neutron star radius $R$,
the envelope can be considered
as effectively flat and the functions $\Phi$ and $\Lambda$ can be replaced by
constants, $2\Phi \approx - 2\Lambda \approx \ln(1-2GM/Rc^2)$. In this
approximation (see \citealt{GPE83}) the hydrostatic equilibrium and heat
diffusion equations can be written as
\beq
   \frac{\dd P}{\dd z}=g_\mathrm{s} \rho,
\quad
   \kappa \frac{\dd T}{\dd z}= F_T
\label{e:flat}
\eeq
where $g_\mathrm{s} = \mathrm{e}^{\Lambda} GM/R^2$ is the surface
gravitational acceleration and $z = \mathrm{e}^{\Lambda} (R-r)$ is the proper
depth.

The system  of equations \eqref{e:SysEq} together with the equation
of state (EOS) and the heat transport equation \eqref{e:HeatEq}
constitute the full set of equations required for calculating the
diffusively equilibrated configuration of the envelope. The
integration is carried out from the atmosphere (with an effective
temperature $\Ts$) to $\rho=\rhob$. This gives the distribution of
all physical quantities (particularly, $T$, $P$, $n_\alpha$) within
the heat blanketing envelope; then we have $\Tb=T(\rhob)$, and
construct the required $\Tb-\Ts$ relation.

For the EOS, we use analytical approximations described in
\citet{PC10}.\footnote{The corresponding Fortran code is available
at \url{http://www.ioffe.ru/astro/EIP/}} The thermal conductivity
$\kappa$ is calculated as the sum of the electron conductivity
$\kappa_\mathrm{e}$ and the photon conductivity
$\kappa_\mathrm{ph}=16\sigma_\mathrm{SB}T^3/3\rho K_\mathrm{rad}$,
where $K_\mathrm{rad}$ is the radiative opacity. For the latter, we
use the Rosseland mean opacities provided either by the Opacity
Library (OPAL, \citealt*{OPAL})\footnote{Available through the MESA
project (\citealt{MESA3} and references therein) at
\url{http://mesa.sourceforge.net/index.html}} or by the Opacity
Project (OP, \citealt{OP} and references therein)\footnote{Available
at \url{http://opacities.osc.edu/rmos.shtml}}. We have checked that
the differences between the OPAL and OP opacities are negligible for
the conditions of our interest. We have performed interpolation
across the radiative opacity tables and extrapolation outside their
ranges in the same way as in \citet{PCY97}. The electron thermal
conductivities $\kappa_\mathrm{e}$ have been calculated using the
approximations described in Appendix~A of \citet{PPP15} (see
references therein for details).\footnote{The corresponding Fortran
code is available at \url{http://www.ioffe.ru/astro/conduct/}}
Typically, photon conduction dominates
($\kappa_\mathrm{ph}>\kappa_\mathrm{e}$) in  the outermost
nondegenerate neutron star layers, whereas electron conduction
dominates in deeper, moderately or strongly degenerate layers
(\citealt{GPE83}).

Equations \eqref{e:SysEq} are analogous to the chemical equilibrium
equations of \citet{CB10}. The difference is in the presence of the
$\bm{\nabla} T$ term in equations \eqref{e:GenForce} and
\eqref{e:CorrGrad}.

\section{Overall description of models}
\label{sec:EnvModel}

We have modeled a number of heat blanketing envelopes composed of
$^1$H -- $^4$He, or $^4$He -- $^{12}$C or $^{12}$C -- $^{56}$Fe
mixtures. Real envelopes can naturally contain other ions; we have
chosen these three BIMs  as important illustrative examples. The
calculations have been performed for the surface gravity
$g_\mathrm{s0}=2.4271\times10^{14}$ cm s$^{-2}$, which corresponds
to the `canonical' neutron star model with the mass $M=1.4$ M$\odot$
and radius $R=10$ km. For two realistic EOS models of neutron star
matter, APR \citep*{APR98} or BSk21
(\citealt*{Goriely_etal10}; \citealt{Potekhin_etal13},
and references therein),
this surface gravity  corresponds to neutron stars with $M=1.73$
M$\odot$ and $R=11.3$ km or with $M=2.00$ M$\odot$ and $R=12.3$ km,
respectively. In the adopted locally flat approximation, the structure
of the envelope will not depend on $M$ and $R$ separately,
but only on the surface gravity $g_\mathrm{s}$.
Such models of heat blankets are self-similar.
It is sufficient to build a model for one value of $g_\mathrm{s}$;
it can be immediately rescaled for another $g_\mathrm{s}$
\citep{GPE83}; also see Appendix \ref{sec:Fits} and equation \eqref{e:EffectiveRho}
below.

It is natural that all our calculations of diffusively equilibrated
envelopes demonstrate stratification of elements. One always has H
on top of He in H -- He envelopes; He on top of C in He -- C
envelopes; and C on top of Fe in C -- Fe ones. Therefore, any
envelope contains an upper layer which mainly consists of lighter
ions; a bottom layer mostly composed of heavier ions; and a
transition layer which is essentially a BIM. The width of the
transition layer is variable (as discussed below).

As far as the ion separation is concerned, the three BIMs of our study
are different. In the H -- He and C -- Fe envelopes the `molecular
weights' $Z_j/A_j$ of ions $j=1$ and 2 are different, and the
separation is mainly gravitational. In the He -- C envelopes the
`molecular weights' are almost equal. Therefore, the separation is
produced by weaker Coulomb forces; the gravitational separation due to
the nuclear mass defects is still much weaker in this case -- see
\citet{CB10,BY13}.

To analyze the results we need a parameter which would characterize
the position of the intermediate layer and the mass $\Delta M$ of
lighter nuclei in the heat blanketing envelope. It is instructive to
introduce the effective transition density $\rho^*$ and pressure
$P^*$ as the density and pressure at such an (artificial) surface
that the total mass $\Delta M$ contained in the outer shell at $P <
P^*$ would be equal to the actual total mass of the lighter ion
species in the absence of diffusive mixing (as if for exact
two-shell structure). In the approximation that all the pressure is
provided by degenerate electrons, one has (e.g.
\citealt{GPE83,PCY97,Ofengeim_etal15})
\begin{align}
\begin{split}
  \frac{\Delta M}{M} = {} &
  \frac{0.838}{g_\mathrm{s14}^2}\,\frac{P^*}{10^{34}
  \textrm{ dyn cm}^{-2}} = \frac{1.510 \times
  10^{-11}}{g_\mathrm{s14}^2} \\
  & \times
  \Bigg\{\xi(\rho^*) \sqrt{1+\xi(\rho^*)^2}
  \left[ \frac{2}{3}\,\xi(\rho^*)^2 - 1 \right] \\
 & +\ln \left[ \xi(\rho^*) + \sqrt{1+\xi(\rho^*)^2}\, \right]
 \Bigg\},
\end{split}
\label{e:EffectiveRho}
\end{align}
where $g_\mathrm{s14}$ is the
surface gravity in units of 10$^{14}$ cm s$^{-2}$,
\begin{equation}
   \xi(\rho) = 0.01009 \, (\rho{Z}/{A})^{{1}/{3}},
\label{e:RelParam}
\end{equation}
is the dimensionless electron relativity parameter (where
$\rho$ is meant to be measured in \gcc), while
$Z$ and $A$ are, respectively, the charge and mass numbers of
lighter ions. Thus we characterize $\Delta M$ by $\rho^*$.

The solution of equation \eqref{e:EffectiveRho} with respect to $\xi$ gives
us the effective transition density $\rho^*$. Starting from an
arbitrary fixed value of $x_1=n_1/n$  near
the surface, we integrate
the system of equations \eqref{e:SysEq}, \eqref{e:CorrGrad} and
\eqref{e:flat} inside the heat blanketing envelope and
obtain different profiles of ion densities $n_j(z)$
($j=1,2$), which correspond to different $\Delta M$ and $\rho^*$. Note
that for  small enough $\rho^*$ the electron degeneracy
can be removed. In such cases  equation
\eqref{e:EffectiveRho} presents just a formal definition
of $\rho^*$ through $\Delta M$;
$\rho^*$ acquires clear meaning of the characteristic transition density if it
belongs to the domain of degenerate electrons.

\section{Parameters of models and their ranges}
\label{sec:ModelParam}

\begin{figure*}
\centering
\includegraphics[height=8.0cm,keepaspectratio=true,clip=true,trim=0.05cm 0.15cm 0.5cm 1cm]{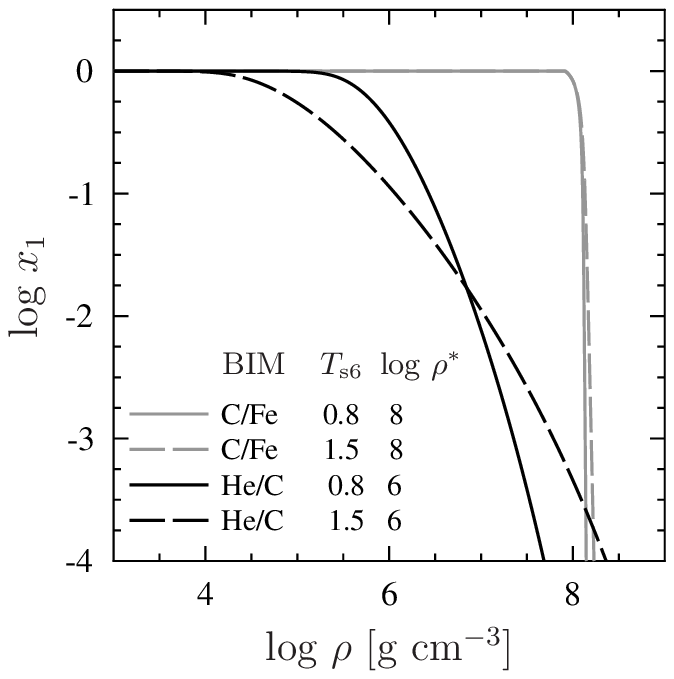}%
\includegraphics[height=8.0cm,keepaspectratio=true,clip=true,trim=0.05cm 0.15cm 0.5cm 1cm]{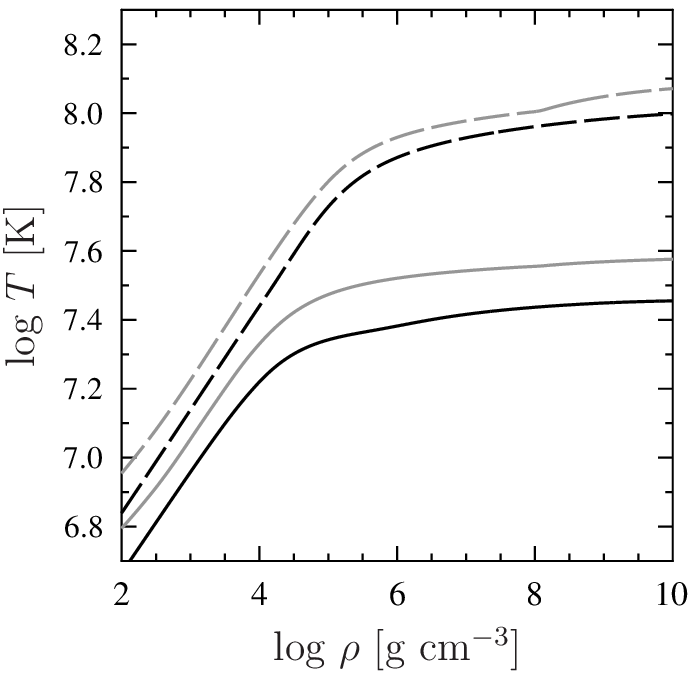}
\caption{Number fractions of lighter elements versus $\rho$
(left-hand panel) and $T(\rho)$ dependence (right-hand panel) in He --
C and C -- Fe heat blanketing envelopes of a `canonical' neutron
star ($M=$ 1.4\,M$\odot$; $R=10$ km). Curves are calculated for
$T_\mathrm{s6}=\Ts/10^6$ K = 0.8 and 1.5 at $\rho^*=10^6$ \gcc{}  (He --
C; black lines) and $10^8$ \gcc{}   (C -- Fe; grey lines). }
\label{fig:start}
\end{figure*}

\begin{figure}
\centering
\includegraphics[viewport = 0 10 225 200,width=0.53\textwidth]{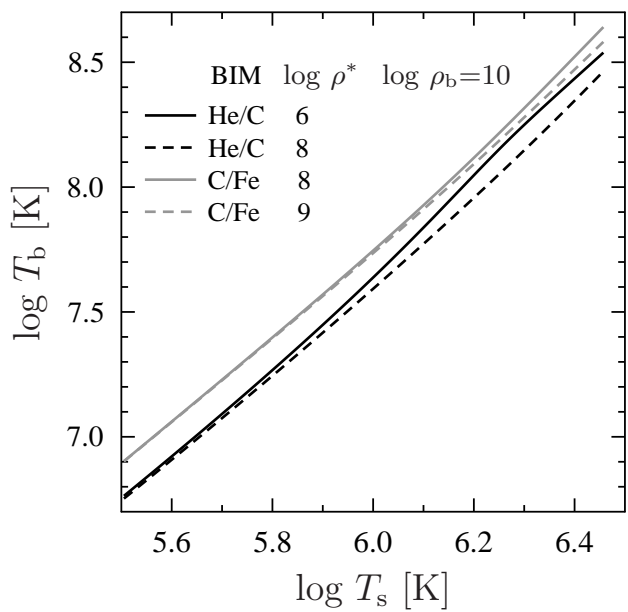}
\caption{$\Tb-\Ts$ relations in He -- C (black lines) and C -- Fe
(grey lines) heat blanketing envelopes of a `canonical' neutron star with
$\rho^*=10^6$ and $10^8$ \gcc{}   for He -- C envelopes (solid and dashed curves,
respectively) and with $\rho^*=10^8$ and $10^9$ \gcc{}   for C -- Fe envelopes
(solid and dashed curves, respectively); $\rhob=10^{10}$ \gcc.
See text for details.}
\label{fig:starta}
\end{figure}

After fixing the surface gravity $g_\mathrm{s}$, our models of
heat blanketing envelopes are characterized by a composition (H --
He, He -- C, or C -- Fe), an effective surface temperature $\Ts$, an
amount of lighter ions in the envelope (specified by $\rho^*$ or
$\Delta M$) and by a density $\rhob$ at the envelope bottom.  The
input parameters are naturally restricted (see, e.g.,
\citealt{PCY97} and references therein). In particular, at high $T$
and/or $\rho$ hydrogen transforms into helium (due to thermo- or
pycno-nuclear burning and beta captures; very roughly, this happens
at $T \gtrsim 4 \times 10^7$ K and/or $\rho \gtrsim 10^7$ \gcc).
Then helium transforms into carbon (at $T \gtrsim 10^8$ K and/or
$\rho \gtrsim 10^9$ \gcc), and carbon transforms into heavier
elements (at $T \gtrsim 10^9$ K and/or $\rho \gtrsim 10^{10}$
\gcc). Another restriction is that $\rho^* \lesssim \rhob$;
otherwise, the heat blanketing envelope is essentially one-component
(consists of lighter ions). The mass $\Delta M$ cannot be smaller
than the mass of the atmosphere (that is typically $\sim
10^{-18}-10^{-16}$ $\Msun$).

\begin{figure*}
\centering
\includegraphics[height=8.0cm,keepaspectratio=true,clip=true,trim=0.05cm 0.15cm 0.5cm 1cm]{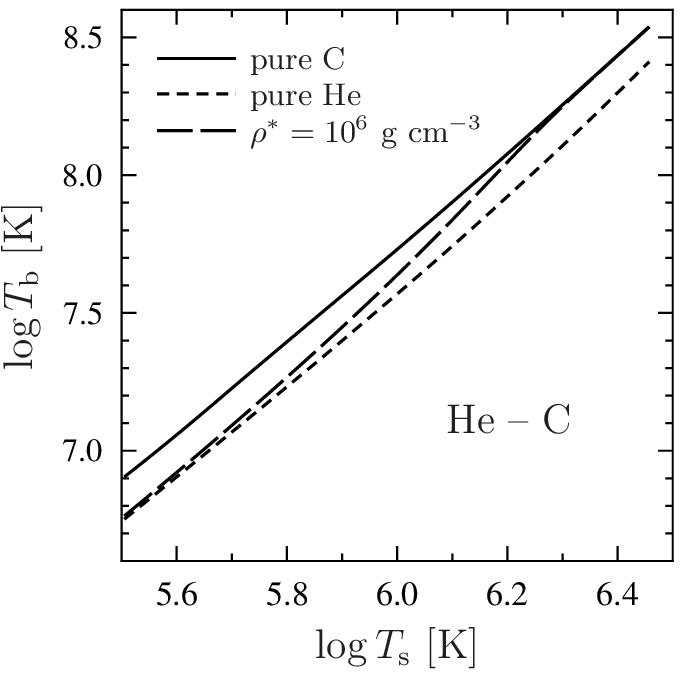}%
\includegraphics[height=8.0cm,keepaspectratio=true,clip=true,trim=0.05cm 0.15cm 0.5cm 1cm]{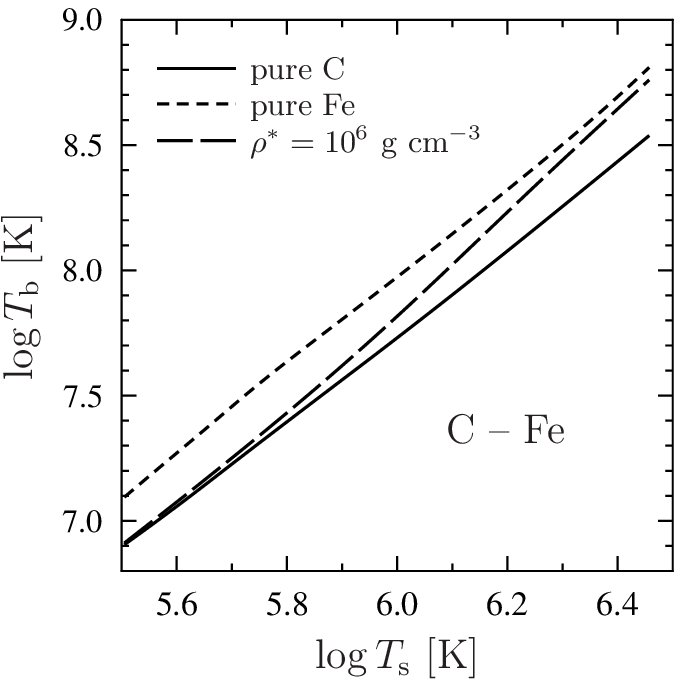}
\caption{Internal temperature $\Tb$ calculated assuming
$\rho^*=10^6$ \gcc{}   and $\rhob = 10^{10}$ g cm$^{-3}$ for a
`canonical' neutron star as a function of the surface temperature
$\Ts$ for He  / C (left-hand panel) and C / Fe (right-hand panel)
pure elements and mixtures. Solid curve refers to pure C on both
panels. Short-dashed curve is for pure He on the left-hand
panel and to pure Fe on the right-hand one. Long-dashed curve refers
to the He -- C mixture on the left and to the C -- Fe mixture on the
right. See text for details.}
\label{fig:Tb-Ts}
\end{figure*}

A choice of $\rhob$ deserves special comments. The introduction of
$\rhob$ accelerates numerical simulations of thermal evolution of
neutron stars. One can use an obtained $\Tb-\Ts$ relation to
simulate the temperature distribution within the star (at
$\rho>\rhob$) taking $T=\Tb$ as a boundary condition. However,
$\Tb-\Ts$ relations are calculated in a stationary approximation.
Therefore, such a boundary condition is valid as long as time
variations of $T$ within the heat blanketing envelope are slower
than typical time $t_\mathrm{d}$ of thermal diffusion through this
envelope. Simple estimates of $t_\mathrm{d}$ for an iron heat
blanketing envelope of a `canonical' neutron star at $\Ts=1$ MK give
$t_\mathrm{d} \sim 1$ yr for $\rhob=10^{10}$ \gcc. With this $\rhob$
one cannot model variations of $\Ts \sim 1$ MK shorter than one
year. Moving $\rhob$ closer to the surface, $\rhob \to 10^8$ \gcc,
one comes to $t_\mathrm{d} \sim 1$ d, which would allow one to simulate
much shorter time variations of $\Ts$ with the cooling code (but the
code could become less efficient). We present the results for
different $\rhob$ which should be helpful for solving different
problems of thermal evolution of neutron stars.

We have constructed many models of heat blanketing envelopes with
different parameters. The effective surface temperature has been
varied from $\Ts \sim 0.3$ MK to $\Ts \sim 3$ MK which is a typical
range of $\Ts$ measured for cooling isolated neutron stars (see
\citealt{Vigano_etal13} and references therein\footnote{A table of
observed characteristics of thermally emitting neutron stars is
available at \url{http://www.neutronstarcooling.info/}.}). For the H
-- He envelopes we have considered $\rhob=10^8$ and $10^9$ \gcc, and
varied $\rho^*$ up to $\sim 10^7$~\gcc. For the He -- C and C -- Fe
envelopes we have taken $\rhob=10^8$, $10^9$ and $10^{10}$ \gcc. In
case of the He -- C envelopes we have varied $\rho^*$ up to $\sim
10^8$ \gcc, and for the C -- Fe envelopes up to $10^{9}$ \gcc. We
have mainly limited our calculations to those cases in which $T(\rho)$
in the envelope is lower than characteristic temperature of nuclear
transformations (see above).

\section{Results for diffusively equilibrated envelopes}
\label{sec:EquilibResults}

\begin{figure*}
\centering
\includegraphics[height=8.0cm,keepaspectratio=true,clip=true,trim=0.00cm 0.15cm 0.5cm 1cm]{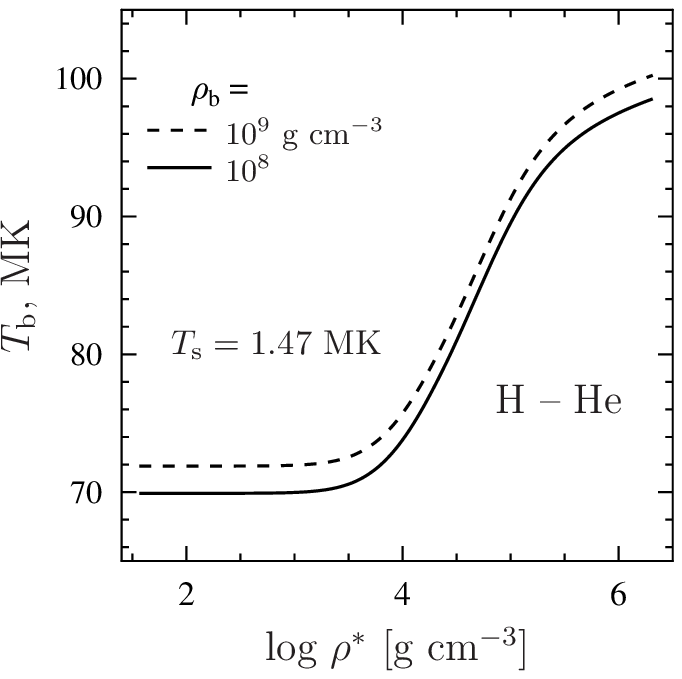}%
\includegraphics[height=8.0cm,keepaspectratio=true,clip=true,trim=0.00cm 0.15cm 0.5cm 1cm]{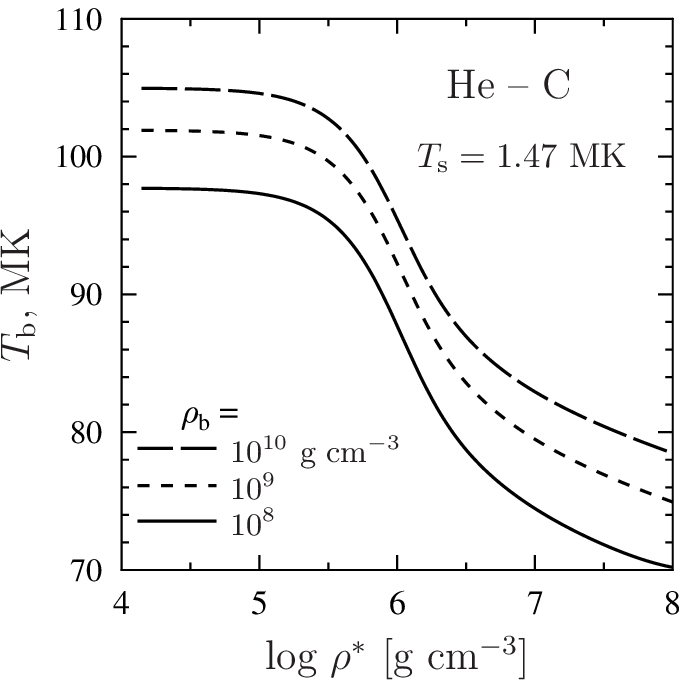}
\caption{Internal temperature $\Tb$ versus $\rho^*$ for a
`canonical' neutron star with a H -- He heat blanketing envelope
extended to $\rhob=10^8$ or $10^{9}$ \gcc{}  (left-hand panel) and
with a He -- C envelope extended to $\rhob=10^8$, $10^9$ or
$10^{10}$ \gcc{}  (right-hand panel). The surface temperature is
$\Ts=1.47$ MK. One can see the transition from the case of purely
heavy ions (low $\rho^*$) to purely light ions (high $\rho^*$). See
text for details.}
\label{fig:Tb-Rhostar}
\end{figure*}

\begin{figure}
\centering
\includegraphics[width=\columnwidth]{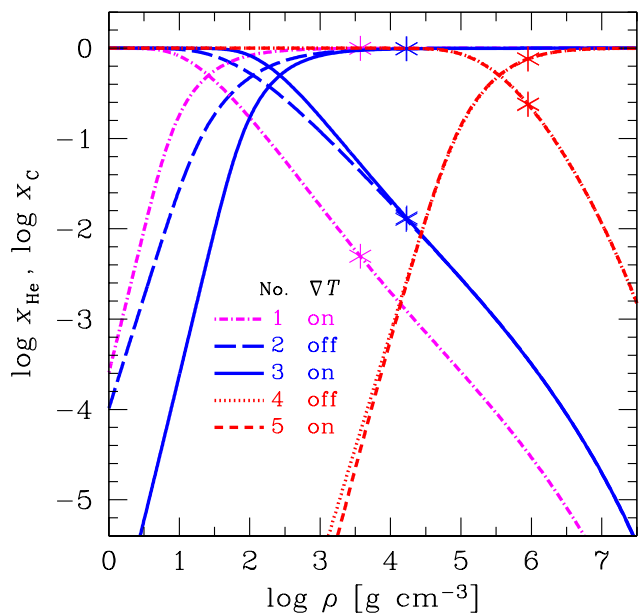}
\caption{Profiles of the helium (decreasing)
and carbon (increasing) number fractions as functions of density
$\rho$ for five models of He -- C envelopes of a `canonical' neutron
star with $\Ts=1.1$ MK. Different models 1\,--\,5  are shown by different line
styles (see text for details). They either include or exclude the $\bm{\nabla}T$ term in equation
(\ref{e:GenForce}), as indicated in the legend. An asterisk on each
curve marks the effective transition density $\rho^*$.}
\label{fig:Diff}
\end{figure}

Fig.{} \ref{fig:start} illustrates the distribution of ions and the
temperature profiles $T(\rho)$ in the He -- C and C -- Fe
envelopes with $\rhob=10^{10}$ \gcc. Calculations are performed
for two surface temperatures, $\Ts=$ 0.8 and 1.5 MK (solid and dashed
lines, respectively). The total amount of lighter ions is fixed to
$\rho^*=10^6$ \gcc{}  for the He -- C envelope (black lines), and to
$10^8$ \gcc{}  for the C -- Fe one (grey lines). Accordingly, the
transition layer from lighter ions to heavier ones for the C -- Fe
envelope lies deeper. The assumed $\rho^*$ in the He -- C envelope
corresponds to the geometrical depth $z^* \approx 3$ m, and the
bottom depth of the envelope is $z_\mathrm{b}\approx 161$ m; for the C
-- Fe envelope, we have $z^* \approx 28$ m and $z_\mathrm{b} \approx
145$ m.

The left-hand panel of Fig.{} \ref{fig:start} demonstrates the
density dependence of the number fraction $x_1$ of lighter ions (He
for He -- C; C for C -- Fe).  One can observe different profiles
$x_1(\rho)$ for the He -- C and C -- Fe envelopes. Characteristic
relative width $\delta \rho/\rho^*$ of the transition layer in the
He -- C envelope is typically more than ten times larger than in the
C -- Fe envelope. This results from much weaker (Coulomb) separation
in the He -- C mixture. If the separation of ions is gravitational
(as in C -- Fe or H -- He BIMs) a transition from lighter to heavier
ions in diffusive equilibrium is rather sharp, but in case of
Coulomb separation (He -- C) it is broad (similar conclusion has
been made by \citealt{CB10}). There appears a tail of He ions at
densities much larger than $\rho^*$; these ions constitute a
noticeable fraction of the total He mass, $\Delta M$. Of course,
similar tail exists also in the C -- Fe mixture, but it is much less
pronounced. When $\Ts$ decreases, the envelopes become colder and
the transition layers narrower.

The right-hand panel of Fig.{} \ref{fig:start} shows the temperature
$T$ versus density in the same envelopes. Because the He -- C
envelope consists of lighter ions, it is overall more heat
transparent, than the C -- Fe envelope, and has a lower $T(\rho)$
for the same $\Ts$. For the densities close to $10^{10}$ \gcc{}  the
thermal conductivity becomes so high that the temperature $T(\rho)$
tends to saturate reaching the temperature of nearly isothermal
matter behind the heat blanketing envelope (\citealt{GPE83,PCY97}).

Fig.{} \ref{fig:starta} displays the $\Tb-\Ts$ relations calculated
for the He -- C and C -- Fe envelopes with $\rhob=10^{10}$ \gcc.
In case of the He -- C envelope, we plot $\Tb - \Ts$ at
$\rho^*=10^6$ and $10^8$ \gcc; while for the C -- Fe envelope at
$\rho^*=10^8$ and $10^9$ \gcc. Because the He -- C envelope is
overall more heat transparent, it has a lower $\Tb$ for the same
$\Ts$. By increasing $\rho^*$ we increase the amount of lighter ions
in a given envelope, which also increases the heat transparency (at
sufficiently high $\rho$) and decreases $\Tb$ (at sufficiently high
$\Ts$ at which the main temperature gradient reaches the range of
$\rho \sim \rho^*$).

Fig.{} \ref{fig:Tb-Ts} shows typical $\Tb-\Ts$ relations for the
He -- C (left-hand panel) and C -- Fe (right-hand panel) envelopes.
On each panel we plot $\Tb-\Ts$ for an envelope containing
pure lighter ions (He or C); pure heavier ions (C or Fe); and
a mix appropriate to $\rho^*=10^6$ \gcc. Envelopes of pure
lighter ions are better heat conductors and have lower $\Tb(\Ts)$.
Envelopes of pure heavier ions are better heat insulators and have higher
$\Tb(\Ts)$. Envelopes containing BIMs produce intermediate heat insulation.
Increasing $\rho^*$ varies their insulation from that for heavier ions
to that for lighter ones.

Fig.{} \ref{fig:Tb-Rhostar} demonstrates the dependence of $\Tb$ on
the transition density $\rho^*$ for the H -- He (left-hand panel)
and He -- C (right-hand panel) envelopes. The surface temperature is
fixed to $\Ts=1.47$ MK. The solid lines are calculated assuming
$\rhob=10^8$ \gcc, the short-dashed lines are for $\rhob=10^9$
\gcc{}  and the long-dashed line for the He -- C envelope is for
$\rhob=10^{10}$ \gcc. We do not present similar line for the H --
He envelope because He cannot survive at such high densities (Sect.\
\ref{sec:ModelParam}). Any line exhibits a transition from the regime of
low $\rho^*$, where the amount of lighter ions is small and the
envelope behaves as almost fully composed of heavier ions, to the
regime of high $\rho^*$, where the amount of heavier ions is small
and the envelope behaves as if it consists of lighter ions. The
ranges of intermediate $\rho^*$ in which the binary composition is
really significant are seen to be wide.

Notice the anomalous
behavior of the H -- He BIM. For this BIM, contrary to the He -- C
and C -- Fe BIMs, increasing the amount of lighter (hydrogen) ions
leads to the growth of $\Tb$. This effect has been
overlooked in previous studies (see, e.g., \citealt{PCY97}) which
stated that replacing He with H does not affect $\Tb$. The effect is
mainly because hydrogen has a different mass to charge ratio than
helium and carbon, and also because of low radiative opacities of
helium. A C -- Fe mixture has the same transition `direction' as He
-- C mixture since the mass to charge ratio of iron is not very
different from that of carbon (unlike hydrogen where the difference
is larger).

Fig.{} \ref{fig:Diff} shows the impact of the $\bm{\nabla}T$ term in
equation \eqref{e:GenForce}, or in \eqref{e:CorrGrad}, on the
properties of He -- C envelopes. The figure shows the helium fraction
profile $x_\mathrm{He}(\rho)$ calculated in five cases (curves 1\,--\,5)
for the same surface temperature $\Ts=1.1$ MK. Cases 1, 3, and 5
are calculated with account of the $\bm{\nabla}T$ term, whereas in
cases 2 and 4 this term is neglected (which is equivalent to the
approximation made by \citealt{CB10,BY13}).  The curves 2 and 3 are
computed for the same effective transition density
$\rho^*\approx1.7\times10^4$ \gcc, whereas model 1 has the same trace
amount of carbon with model 2 at the radiative surface, from which we
start the integration [$x_\mathrm{C}(z=0)=2 \times10^{-6}$]. The
latter boundary condition leads to a different accumulated He mass, that
is to different transition density $\rho^*\approx 3.7\times10^3$ \gcc.
However, the differences between the curves 1, 2 and 3 are
insignificant for the $\Tb-\Ts$ relation. The calculated $\Tb$ values
differ by $\lesssim1$ per cent, because the corresponding $\rho^*$ lie
outside the `sensitivity strip' \citep{GPE83} which is the $\rho-T$ domain where the conductivity affects the $\Tb-\Ts$ relation most significantly. At contrast, both
models 4 and 5 have $\rho^*\approx9\times10^5$ \gcc{}  inside the
sensitivity strip, but in this case the $\bm{\nabla}T$ term is less
significant because of stronger degeneracy. As a consequence,
the curves 4 and 5 are very close to each other, so that the $\bm{\nabla}T$ term is
also unimportant for the $\Tb-\Ts$ relation (the difference in $\Tb$
is again within 1 per cent).

We note that in the cases 1\,--\,3 the He abundance is quite low,
$x_\mathrm{He}\lesssim0.01$, at the transition density $\rho^*$. This
reflects the fact that in these three cases the layer with high He
abundance is mostly nondegenerate, but a considerable fraction of the
total He mass is supplied by a diffusive tail in the deeper degenerate
layers of the envelope.

Our calculations show that the $\bm{\nabla}T$ term significantly
affects the ion fractions if the layer, where the ion Coulomb coupling
is moderate (neither weak nor strong), is close to the layer, where a
transition from lighter to heavier ions takes place. Such situations
may occur at sufficiently high $\Ts$ in the outer layers
($\rho\lesssim 10^7$ \gcc) of the envelopes composed of sufficiently
light elements like hydrogen, helium or carbon. Even in these cases
the $\Tb-\Ts$ relations, the pressure and total density profiles are
affected much  weaker by the $\bm{\nabla}T$ term. Moreover, in the
limit of strong  Coulomb coupling (described, e.g. in \citealt{BY13})
the $\bm{\nabla}T$ term vanishes completely and non-isothermal
calculations coincide exactly with isothermal ones (as long as we do
not take into account thermal diffusion).

\begin{figure}
\centering
\includegraphics[viewport = 0 10 225 200,width=0.53\textwidth]{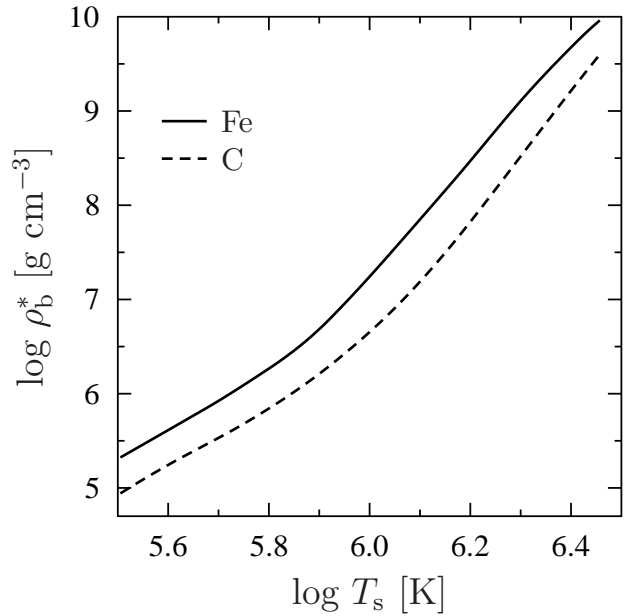}
\caption{Effective density $\rhob^*$ at the bottom of the heat
blanketing envelope composed either of pure Fe (solid line) or of
pure C (dashed line) as a function of $\Ts$ for a `canonical'
neutron star. See text for details. }
\label{fig:Rho_b-Ts}
\end{figure}

Finally, Fig.{} \ref{fig:Rho_b-Ts} illustrates another important feature
of heat blanketing envelopes which is not related directly to their
multicomponent structure. Specifically, it concerns the meaning of
$\rhob$. If one integrates the equations of thermal structure for a
heat blanketing envelope from the surface to the bottom
($\rho=\rhob$), one often obtains (e.g. Fig.{} \ref{fig:starta}) that
the growth of $T(\rho)$ nearly saturates at some
$\rho=\rhob^*<\rhob$. This saturation is evidently associated with
the growth of the thermal conductivity within the star. It is
especially pronounced in a cold neutron star manifesting the
appearance of the inner isothermal region $\rho>\rhob^*$ within the
star. In contrast to the density $\rhob$ which is \emph{artificially
assumed}, $\rhob^*$ can be viewed as a \emph{real physical} bottom
density of the heat blanketing envelope. Fig.{} \ref{fig:Rho_b-Ts} shows
this density for a `canonical' neutron star, whose envelope consists
solely either of iron or carbon. In a hot star ($\Ts \sim 3$ MK),
the physical heat blanket is thick (close to the assumed heat
blanket with $\rhob \sim 10^{10}$ \gcc). However when the star
cools, $\rhob^*$ decreases, implying that $\Tb$ is actually
determined by a much thinner `physical' heat insulating layer. For
instance, at $\Ts=1$ MK we have $\rhob^* \lesssim 10^{7}$ \gcc{}  so
that $\Tb$ becomes insensitive to the physics of matter at higher
densities (to the composition of such a matter and to
whether it is liquid or solid). The colder the star, the thinner the
`physical' heat blanket. On the other hand, let us remind that the
blanket can become thick, with $\rhob^*>10^{10}$ \gcc, for
magnetars, as shown by \citet{PCY07}. In a multilayer heat blanketing
envelope it is also possible to encounter a `false physical bottom',
where $T(\rho)$ saturates at certain $\rhob^*$, but resumes
its growth at a larger density when it enters a layer with a higher $Z$.

\section{Non-equilibrium heat blanketing envelopes}
\label{sec:NonEquilib}

\begin{figure*}
\centering
\includegraphics[height=8.0cm,keepaspectratio=true,clip=true,trim=0.05cm 0.15cm 0.5cm 1cm]{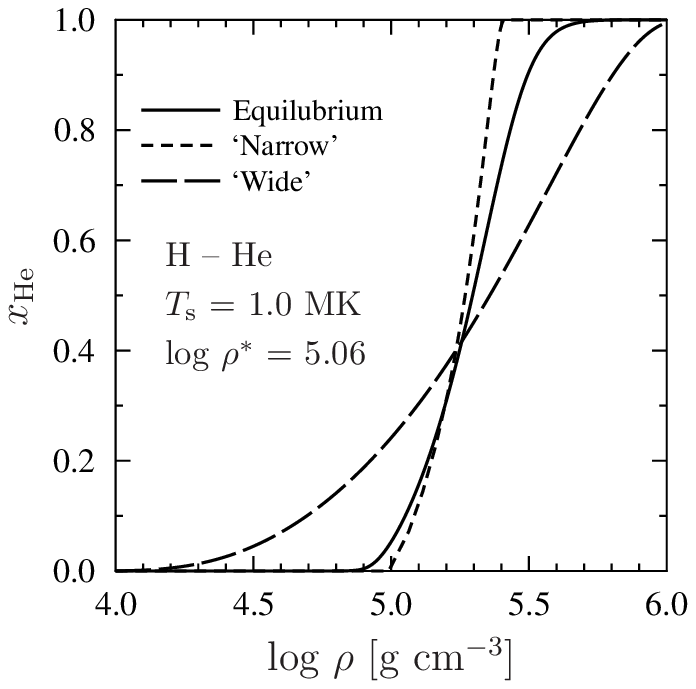}%
\includegraphics[height=8.0cm,keepaspectratio=true,clip=true,trim=0.05cm 0.15cm 0.5cm 1cm]{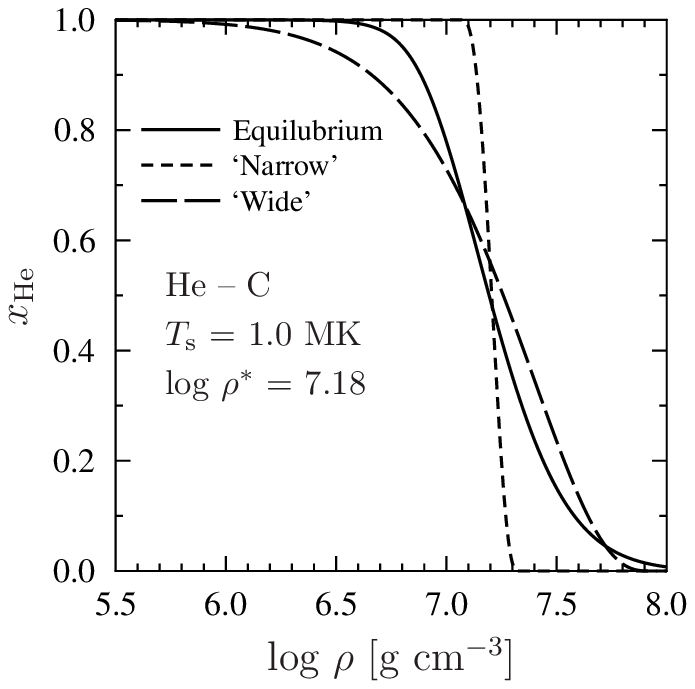}
\caption{Helium number fraction versus density in heat
blanketing envelopes of a `canonical' neutron star with
$\Ts=1$ MK containing H and He (left-hand panel;
$\Delta M=5.09 \times 10^{-14}$ M$\odot$,
$\log \rho^*=5.06$) or He and C (right-hand panel;
$\Delta M=3.04 \times 10^{-11}$ M$\odot$;
$\log \rho^*=7.18$). Solid lines refer to the envelopes in
diffusive equilibrium while short-dashed and long-dashed lines are
for the envelopes out of diffusive equilibrium, with narrower
and wider transition layers, respectively. See text for details.}
\label{fig:x-rho}
\end{figure*}

In addition to diffusively equilibrated heat blanketing envelopes
considered above, we have also studied the envelopes out of
diffusive equilibrium. Since ion diffusion is rather slow (see
below) such envelopes can exist for a long time (being, of course,
in the overall hydrostatic equilibrium). For illustration, we study
them in a quasi-statical approximation, fix the distribution of
ions, $x_j(\rho)$, disregard the diffusive equilibrium and calculate
the structure of the envelopes by integrating equations
(\ref{e:flat}). This is much easier than respect the diffusive
equilibrium.

Some illustrative results are shown in Fig.{} \ref{fig:x-rho}.
On the left-hand panel we present three models of H -- He envelopes,
and on the right-hand panel three models of He -- C envelopes. The
figure shows the profile of the helium number fraction $x_\mathrm{He}$
versus $\rho$ for a `canonical' neutron star. The surface temperature
is fixed to $\Ts=1$ MK for all models. All the three H -- He models
have the same amount of hydrogen ($\log \rho^*=5.06$ [\gcc]) and
all the three He -- C models the same amount of He ($\log \rho^*=7.18$).
The helium fraction decreases with $\rho$ on the left-hand panel
(because He ions are heavier than H) and increases with $\rho$
on the right-hand panel (because He ions are lighter than C ones).
The solid line on each panel corresponds to diffusively equilibrated
envelopes (calculated as described in the previous sections). The
dashed lines are for the envelopes taken to be out of
diffusive equilibrium. The short-dashed lines refer to narrower (than in
diffusive equilibrium) transition layers while the
long-dashed lines refer to wider layers.

It is remarkable, that for all the three He -- C models
we obtain almost the same $\Tb=4.00 \times 10^7$ K
(which we present for $\rhob=10^{10}$ \gcc, as an example).
The same is true for H -- He models.
For instance, assuming $\rhob=10^9$ \gcc{}  we
have $\Tb= 4.64 \times 10^7$ K for the equilibrium and narrower
transition layers and
$\Tb=4.54 \times 10^7$ K for the wider transition layer.
Therefore, the resulting $\Tb-\Ts$ relations seem highly
insensitive to the actual state of the envelope, whether
it is equilibrated or not. These $\Tb-\Ts$ relations are mainly
determined by the mass $\Delta M$ of lighter ions (or, equivalently,
by $\rho^*$). Of course, this statement is true for the envelopes
where the distribution of ions is not too much wider than the
equilibrium one. This is illustrated by a relatively large
deviation from the equilibrium for the wider H -- He distribution;
in this case $\Tb$ becomes slightly different from the equilibrium one.
However, large deviations from equilibrium are expected to
relax at short timescales (days to years, see below).

The insensitivity of $\Tb-\Ts$ relations to number fraction distributions throughout the envelopes also answers the question on thermal  diffusion. Although thermal diffusion
can change the ion fractions, this change would not affect the resulting $\Tb-\Ts$ relation. However, if one is interested in the processes which are sensitive to
number fractions (e.g., diffusive nuclear burning) then thermal diffusion
can be important. We have made order of magnitude estimates of the impact of thermal diffusion
on the diffusion velocity. We have assumed a constant thermal diffusion ratio $k_T=0.1$.
This is a conservative upper limit obtained in our calculations with
the effective potential method described by \citealt{BY14};
real values are smaller. For H -- He mixture ($x_{\mathrm{H}}=x_{\mathrm{He}}=0.5$)
the thermal diffusion correction to the diffusion velocity does not exceed 3 per cent,
while for He -- C mixture ($x_{\mathrm{He}}=x_{\mathrm{C}}=0.5$)
it does not exceed 6 per cent.
This correction has its largest value near the surface where the temperature gradient is big
(see, e.g., the right-hand panel of Fig.{} \ref{fig:start}) and decreases with depth.

Using equation (\ref{e:J-BIM}) and taking typical depth-scales $\Delta z$ of
deviations from diffusive equilibrium in the transition layer,
for the conditions in Fig.{} \ref{fig:x-rho} we can estimate characteristic relative
velocities $V$ of two ion species during diffusive equilibration in that layer and typical
equilibration times $t_\mathrm{eq}\sim \Delta z /V$. For H -- He
envelopes (left-hand panel) we very roughly obtain
$\Delta z\sim$ a few meters, the equilibration velocity
 $V \sim 10^{-4}-10^{-3}$ cm~s$^{-1}$,
and the equilibration time $t_\mathrm{eq} \sim$ one or a few days.
For He -- C envelopes (right-hand panel) we also have
$\Delta z\sim$ a few meters, but the diffusive velocities
$V \sim 10^{-7}-10^{-6}$ cm~s$^{-1}$ are lower,
and $t_\mathrm{eq}\sim$ a few years. The equilibration in the
He -- C envelopes goes much slower because of weaker Coulomb separation
and deeper transition layer. Our example shows that
the He -- C envelopes can be out of diffusive equilibrium for a long time.


\section{Conclusions}
\label{sec:concl}

We have considered two-component heat blanketing envelopes of neutron stars.
These envelopes can be either in diffusive equilibrium or out of it. Our main goal has
been to relate the effective surface temperature of the star, $\Ts$, to
the temperature $\Tb$ at the bottom of the envelope ($\rho=\rhob \sim
10^8 - 10^{10}$ \gcc) and to investigate the sensitivity of this relation
to the distribution of ion species within the envelope.

We have derived general expressions for the diffusive fluxes in
multicomponent non-isothermal gaseous or liquid Coulomb systems of
ions with arbitrary Coulomb coupling taking into account temperature
gradient. In the limit of weakly coupled plasma these expressions
reproduce the classical expressions for diffusion in ideal gas
mixtures. Our new expressions are valid not only for Coulomb systems,
but also for any gaseous or liquid system (diffusion is also available
in solids, e.g. \citealt{Hughto11}, but it is greatly suppressed there
compared to gases  and liquids).

For applications, we have calculated the $\Tb-\Ts$ relations for two
component envelopes (containing H -- He, He -- C, or C -- Fe mixtures).
These envelopes are naturally stratified into three layers.
The outer layer consists predominantly of lighter ions;
the inner layer near the envelope bottom contains mainly heavier ions; and there
is a transition layer of essentially binary mixture in between.
The stratification in the H -- He and C -- Fe envelopes, where two ion species have
different `molecular weights', is mainly gravitational; while in the
He -- C envelopes it is much weaker (Coulombic).
Accordingly, the transition layers in the He -- C envelopes are much wider
than in other envelopes. The $\Tb-\Ts$ relations have been determined for diffusively equilibrated envelopes with different mass $\Delta M$ of lighter ions
(or, equivalently, with different
characteristic densities $\rho^*$ which specify the position of the
transition layer). The results are approximated by analytic expressions
in Appendix \ref{sec:Fits}, which can be used for simulating thermal evolution
of isolated and accreting neutron stars and related phenomena
(e.g., \citealt{CB03,CB04,YP04,CB10,PPP15}).

The most striking result of our analysis is that the $\Tb-\Ts$ relations are
fairly independent of the structure of the transition layer (of its width,
distribution of ions, and of whether it is diffusively equilibrated or
not). These relations depend only on $\Delta M$ (or on
$\rho^*$). This allows us to expect that the fit expressions presented in
Appendix \ref{sec:Fits} can be used not only for diffusively
equilibrated envelopes but also for
a much wider class of envelope models. In particular, this remarkable property
justifies previous studies \citep{PCY97,Yakovlev_etal11}
of heat blanketing envelopes as a sequence of
layers composed of single ion species (e.g., H, He, C, Fe); slow diffusion
of ions does not introduce noticeable changes in $\Tb-\Ts$ relations.
However, nuclear transformations, which can noticeably change $\Delta M$, can affect these
relations indirectly \citep{CB04,CB10}.

Thus, we have confirmed the previous $\Tb-\Ts$ relations and extended
their studies. First of all, we have considered H -- He and
He -- C envelopes, and approximated the appropriate $\Tb-\Ts$ relations
by analytic expressions for different $\rho^*$ and $\rhob $ in Appendix \ref{sec:Fits}.
We have also reconsidered C--Fe envelopes, found good agreement with
previous results \citep{Yakovlev_etal11}, and fitted the $\Tb-\Ts$ relations
(Appendix \ref{sec:Fits}).

It is evident that our two-component envelopes are idealized; real
envelopes may contain much more ion components. However, ion
stratification seems to be rather strong to prevent the appearance
of layers of essentially multicomponent mixtures if the heat
blanketing envelopes contain many ion species. It is likely that
real envelopes have onion-like structure. Let us stress
once more a great difference of gravitational and Coulomb stratifications.
The latter one is much slower so that the ions with the same charge-to-mass
ratio (like He and C) are mixed much easier than other ions, have
much thicker transition layers, and can be out of diffusive equilibrium
for a longer time. They can form much more extended `tails' outside the
transition layer which can affect nuclear burning, thermal conduction and
other processes important for thermal structure and evolution
of neutron stars. Similar stratification features may be important in white dwarfs.

The expressions for the diffusive fluxes combined with the diffusion
coefficients (see, e.g., \citealt{BY14}) allow one not only to calculate the
diffusively equilibrated configurations of heat blanketing envelopes of neutron
stars, but also the equilibration of these configurations with time.

\section*{acknowledgements}

The work of MB was partly supported by the Dynasty
Foundation, and the work of AP by the Russian Foundation for
Basic Research (grant 14-02-00868-a).



\appendix
\section{Data fitting}
\label{sec:Fits}

\renewcommand{\arraystretch}{1.10}
\setlength{\tabcolsep}{4.5pt}
\begin{table*}
\caption{Fit parameters for H -- He mixture}
\centering
\begin{tabular}{l c c c c c c c c c c c c c c}
\toprule
&  $p_1$  &  $p_2$  & $p_3$  &  $p_4$  &  $p_5$  &  $p_6$  & $p_7$  &  $p_8$  &  $p_9$  &  $p_{10}$  & $p_{11}$  &  $p_{12}$  & $p_{13}$  &  $p_{14}$  \\
\midrule
$\log_{10} \rhob  =8.0$  &  3.150  &  1.546  &  0.3225  &  1.132  &  1.621  &  1.083  &  7.734  &  1.894  &  $2.335\!\times\!10^5$   &  7.071  &  5.202  &  10.01  &  2.007  &  0.4703      \\
\hline
\end{tabular}
\label{tab:H-He-FitParam}
\end{table*}
\setlength{\tabcolsep}{6pt}
\renewcommand{\arraystretch}{1.0}
\renewcommand{\arraystretch}{1.10}
\setlength{\tabcolsep}{4.5pt}
\begin{table*}
\caption{Fit parameters for He -- C mixture}
\centering
\begin{tabular}{l c c c c c c c c c c c}
\toprule
&  $p_1$  &  $p_2$  & $p_3$  &  $p_4$  &  $p_5$  &  $p_6$  & $p_7$  &  $p_8$  &  $p_9$  &  $p_{10}$  & $p_{11}$   \\
\midrule
$\log_{10} \rhob =8.0$  &  5.161  &  0.03319  &  1.654  &  3.614  &  0.02933  &  1.652  &   $1.061\!\times\!10^5$  &  1.646  &  3.707  &  4.011  &  1.153    \\
$\log_{10} \rhob =9.0$  &  5.296  &  0.07402  &  1.691  &  3.774  &  0.08210  &  1.712  &   $1.057\!\times\!10^5$  &  1.915  &  3.679  &  3.878  &  1.110    \\
$\log_{10} \rhob =10.0$  &  5.386  &  0.1027  &  1.719  &  3.872  &  0.1344  &  1.759  &   $1.056\!\times\!10^5$  &  1.881  &  3.680  &  3.857  &  1.102    \\
\hline
\end{tabular}
\label{tab:He-C-FitParam}
\end{table*}
\setlength{\tabcolsep}{6pt}
\renewcommand{\arraystretch}{1.0}
\renewcommand{\arraystretch}{1.10}
\setlength{\tabcolsep}{3.6pt}
\begin{table*}
\caption{Fit parameters for C -- Fe mixture}
\centering
\begin{tabular}{l c c c c c c c c c c c c c c}
\toprule
&  $p_1$  &  $p_2$  & $p_3$  &  $p_4$  &  $p_5$  &  $p_6$  & $p_7$  &  $p_8$  &  $p_9$  &  $p_{10}$  & $p_{11}$  &  $p_{12}$  & $p_{13}$  &  $p_{14}$  \\
\midrule
$\log_{10} \rhob =8.0$    &  0.2420  &  0.4844  &  38.35  &  0.8680  &  5.184  &  1.651  & -0.04390  &  0.001929  &  $3.462\!\times\!10^4$   &  2.728  &  4.120  &  2.161  &  2.065  &   0.008442     \\
$\log_{10} \rhob =9.0$    &  0.1929  &  0.4239  &  48.72  &  1.423  &  5.218  &  1.652  & 0.001037  &  0.004236  &  $3.605\!\times\!10^4$   &  2.119  &  4.014  &  1.943  &  1.788  &   0.01758     \\
$\log_{10} \rhob =10.0$  &  0.1686  &  0.3967  &  55.94  &  1.992  &  5.208  &  1.651  &  0.03235  &  0.005417  &  $3.652\!\times\!10^4$   &  1.691  &  3.930  &  2.021  &  1.848  &  0.02567  \\
 \hline
\end{tabular}
\label{tab:C-Fe-FitParam}
\end{table*}
\setlength{\tabcolsep}{6pt}
\renewcommand{\arraystretch}{1.0}
\renewcommand{\arraystretch}{1.10}
\setlength{\tabcolsep}{4.5pt}
\begin{table}
\caption{Fit errors for H -- He, He -- C and C -- Fe mixtures; $\delta_{\mathrm{rms}}$ stands for rms
relative error, while $\delta_{\mathrm{max}}$
is the maximum relative error. Last column
gives the point where the maximum relative error is achieved.}
\centering
\begin{tabular}{l l c c c}
\toprule
 Mixture & log$_{10}\rhob$ & $\delta_{\mathrm{rms}}$   &  $\delta_{\mathrm{max}}$  &  $\left(Y;~\rho^*/\gcc \right)$\\
\midrule
H -- He &  8.0     &  0.0031  &  0.015  &  $\left(2.865, 3.345\!\times\!10^5\right)$        \\
 \hline
He -- C & 8.0     &  0.0036  &  0.011  &  $\left(0.32, 1.245\!\times\!10^3\right)$        \\
He -- C & 9.0     &  0.0036  &  0.011  &  $\left(0.32, 1.657\!\times\!10^3\right)$        \\
He -- C & 10.0    &  0.0035  &  0.010  &  $\left(0.32, 1.245\!\times\!10^3\right)$      \\
 \hline
C -- Fe & 8.0     &  0.0051  &  0.017  &  $\left(2.865, 1.528\!\times\!10^4\right)$        \\
C -- Fe & 9.0     &  0.0048  &  0.015  &  $\left(0.4259, 1.772\!\times\!10^3\right)$        \\
C -- Fe & 10.0   &  0.0047  &  0.014  &  $\left(0.3872, 1.637\!\times\!10^3\right)$      \\
 \hline
\end{tabular}
\label{tab:FitErr}
\end{table}
\setlength{\tabcolsep}{6pt}
\renewcommand{\arraystretch}{1.0}

We have constructed accurate fits to all computed $\Tb(\Ts, \rho^*)$ data.
These fits have the same general form, but the details depend on a particular
mixture. The general form reads
\begin{align}
\begin{split}
   \Tb \left(Y, \rho^* \right) =
     10^7~\mathrm{K}\, \times & \left\{^{}f_4(Y) +
    \left[f_1 (Y)  - f_4(Y) \right] \right. \\
\times & \left. \left[ 1 + \left( {\rho^*}/{f_2(Y)}\right)^{f_3(Y)} \right]^{f_5(Y)} \right\},
\end{split}
\label{e:GenFit}
\end{align}
where functions $f_1,\ldots, f_5$ are specific for each mixture and
$Y=(\Ts / 1~\mathrm{MK}) \left({g_{s0}}/{g_\mathrm{s}} \right)^{{1}/{4}}$.
The latter relation provides
scaling of $\Tb$ with $g_\mathrm{s}$ \citep{GPE83}, making the
fits valid for any $g_\mathrm{s}$;
$g_\mathrm{s0}=2.4271\times10^{14}$ cm s$^{-2}$ is the value of
$g_\mathrm{s}$ used in our computations; $\Ts$ is the surface temperature
for a star with the surface gravity $g_\mathrm{s}$; $Y$ has meaning of
the surface temperature expressed in MK
for the star with the surface gravity $g_\mathrm{s0}$.

For the H -- He envelopes,
\begin{align}
\begin{split}
   &f_1(Y) = p_1 Y^{p_2} \sqrt{1+p_3 Y^{p_4}}, \\
   &f_4(Y) = p_5 Y^{p_6} \sqrt{1+p_7 Y^{p_8}},\\
   &f_2(Y) = \frac{p_9 Y^{p_{10}}}{\left(1-p_{11} Y + p_{12}Y^2\right)^2}, \\
   &f_3(Y) = p_{13} Y^{-p_{14}} ,\quad
      f_5(Y) = -0.3.
\end{split}
\label{e:H-He-Fit}
\end{align}
The values of the fit parameters are presented in Table
\ref{tab:H-He-FitParam},
and the fit errors are in Table \ref{tab:FitErr}.

For the He -- C envelopes,
\begin{align}
\begin{split}
   &f_1(Y) = p_1 Y^{p_2 \log_{10} Y + p_3}, \quad
      f_4(Y) =  p_4 Y^{p_5 \log_{10} Y + p_6}, \\
   &f_2(Y) = p_7 Y^{p_8 \left(\log_{10} Y \right)^2 + p_9}, \\
   &f_3(Y) = p_{10} \sqrt{\frac{Y}{Y^2+p_{11}^2}} ,\quad
     f_5(Y) = -0.2.
\end{split}
\label{e:He-C-Fit}
\end{align}
The fit parameters and errors are given in Tables \ref{tab:He-C-FitParam} and
\ref{tab:FitErr}, respectively.

Finally, for the C -- Fe envelopes,
\begin{align}
\begin{split}
   &f_1(Y) = p_1 Y^{-p_2}\left(p_3 Y^2 + p_4 Y^4 - 1 \right), \\
   &f_4(Y) = p_5 Y^{p_6}\left(1+p_7 Y^2 -p_8 Y^4 \right), \\
   &f_2(Y) = p_9 Y^{p_{11}-p_{10} \left( \log_{10} Y \right)^2}, \\
   &f_3(Y) = p_{12} \sqrt{\frac{1}{Y^2+p_{13}^2}}\left(1-p_{14}Y^2\right) , \
     f_5(Y) = -0.4.
\end{split}
\label{e:C-Fe-Fit}
\end{align}
The fit parameters are given in Table \ref{tab:C-Fe-FitParam} and the fit errors are listed
in Table \ref{tab:FitErr}.

For each mixture, all parameters have been computed via two-dimensional
fitting procedure; all $(Y, \rho^*)$ points have been fitted
simultaneously. The target function to minimize has been the relative root
mean square error (rms error). The range of fitted data is as follows. For all
mixtures $Y$ spans from 0.32 to $\approx 2.865$ in uniform mesh in logarithmic
scale, 24 points in total. The range of mesh points of
$\rho^*$ differs from mixture to mixture. For H
-- He envelopes, $\rho^*$ spans from $\approx 19.42$ \gcc{}  to
$\approx 3.737\times 10^6$ \gcc{}  and forms
a nonuniform mesh of 41 points. The nonuniformity cannot be avoided;
the internal mesh used in computations is uniform in logarithmic
scale in both $Y$ and $\rho_{\mathrm{int}}$, but when calculating
$\rho_{\mathrm{int}} \to \Delta M \to \rho^*$ the mesh in $\rho^*$ becomes
nonuniform and $Y$-dependent. For He -- C envelopes, $\rho^*$ spans
from $\approx 280.5$ \gcc{}  to $10^8$ \gcc{}  (maximum span, see below)
and also forms a nonuniform mesh. As helium cannot exist
at densities higher than
$10^9$ g cm$^{-3}$ (Sect.{} \ref{sec:ModelParam}), all data points
with $\rho^* > 10^8$ \gcc{}  have been excluded from fitting.
Thus, for different $Y$ values there is different number of points in
$\rho^*$. For C -- Fe envelopes $\rho^*$
spans from $\approx 1459$ \gcc{}  to $\approx 10^9$ \gcc{}  and
forms nonuniform mesh, 40 points in total in $\rho^*$ axis.

Note that for all mixtures the computed data form non-rectangular domains in
the $(Y, \rho^*)$-plane. The domains have a shape of quadrilateral with two parallel
sides (corresponding to $Y$ axis). The above-mentioned range of $\rho^*$ is the
maximum span (i.e. it does not correspond to any $Y$ value; for each $Y$ value
the actual span is smaller and depends on $Y$). For C -- Fe mixture the domain is
close to rectangular one. Nevertheless,
this does not limit the usage of the presented fits. Due to their form
\eqref{e:GenFit}, which reproduces a smooth transition from the
temperature determined by $f_1$ to the temperature determined by $f_4$, they
can be safely extrapolated in $\rho^*$ axis beyond their original domain. On the other hand, the extrapolation in $Y$-direction is not possible (however, if needed, it could be easily constructed based on the presented fits).

\bsp
\label{lastpage}
\end{document}